\documentclass[aip,jcp,superscriptaddress,amsmath,amssymb,twocolumn,reprint]{revtex4-1}

\usepackage[version=3]{mhchem} % Formula subscripts using \ce{}
\usepackage{graphicx}
\usepackage{amsmath}
\usepackage{amssymb}
\usepackage{nicefrac}
\usepackage{booktabs}
\usepackage{mwe}
\usepackage{array}
\newcolumntype{m}{>{$} c <{$}}
\usepackage{color}

\def\rv{{\bf r}}
\def\fv{{\bf f}}
\def\dd{\mathrm{d}}

\def\beq{\begin{equation}}
\def\eeq{\end{equation}}

\def\half{\frac{1}{2}}

\begin{document}     

\author{Stefan Vuckovic}
\affiliation
{Department of Theoretical Chemistry and Amsterdam Center for Multiscale Modeling, FEW, Vrije Universiteit, De Boelelaan 1083, 1081HV Amsterdam, The Netherlands}
\author{Mel Levy}
\affiliation
{Department of Chemistry and Quantum Theory Group, Tulane University, New Orleans, Louisiana 70118, USA}
\author{Paola Gori-Giorgi}
\affiliation
{Department of Theoretical Chemistry and Amsterdam Center for Multiscale Modeling, FEW, Vrije Universiteit, De Boelelaan 1083, 1081HV Amsterdam, The Netherlands}
%\email{}

\title{Augmented potential, energy densities, and virial relations in the weak- and strong-interaction limits of DFT}
\begin{abstract}

The augmented potential introduced by Levy and Zahariev [Phys. Rev. Lett. 113, 113002 (2014)] is shifted with respect to the standard exchange-correlation potential of Kohn-Sham density functional theory by a density-dependent constant that makes the total energy become equal to the sum of the occupied orbital energies. In this work we analyze several features of this approach, focusing on the limit of infinite coupling strength, and studying the shift and the corresponding energy density at different correlation regimes. We also study other definitions of the energy density in relation to local interpolations along the adiabatic connection, present and discuss coordinate scaling properties, and use the shift to analyze the classical jellium and uniform gas models. 

\end{abstract}

\maketitle

%%%%%%%%%%%%%%%%%%%%%%%%%%%%%%%%%%%%%%%%%%%%%%%%%%%%%%%%%%%%%%%%%%%%%
%% The manuscript does not need to include \maketitle, which is
%% executed automatically.  The document should begin with an
%% abstract, if appropriate.  If one is given and should not be, the
%% contents will be gobbled.
%%%%%%%%%%%%%%%%%%%%%%%%%%%%%%%%%%%%%%%%%%%%%%%%%%%%%%%%%%%%%%%%%%%%%
\section{Introduction}
\label{sec_introedens}

The most common successes and failures of the use of Kohn-Sham Density Functional Theory (KS DFT)\cite{KohSha-PR-65} depend on the approximate exchange-correlation (xc) functionals. This is the key quantity in KS DFT and, as its exact form is computationally intractable for systems with more than a few particles,\cite{ColSav-JCP-99,TeaCorHel-JCP-09} the xc functional must be approximated. Density functional approximations (DFA) offer a variety of models that can be more or less expensive to compute and more or less accurate.\cite{Bur-JCP-12,Bec-JCP-14,ZhaTru-ACR-08,SunRemZhaSunRuzPenYanZenPauWagWuKlePer-NATC-16,ErhBleGor-PRL-16}  One of the main problems of most of present DFAs is that they have a bias towards the weak correlation regime. Attempting to avoid this bias, it has been  proposed to construct the xc functional via a {\it local interpolation along the adiabatic connection} between the weak and strong coupling limits.\cite{locpaper,ZhoBahErn-JCP-15,BahZhoErn-JCP-16,VucIroWagTeaGor-PCCP-17,VucGor-JPCL-17} It has been shown that this local approach is generally more accurate and more amenable to the construction of size-consistent approximations than their global\cite{Ern-CPL-96,SeiPerLev-PRA-99,SanCohYan-JCP-06,TeaCorHel-JCP-10} (i.e., integrated over all space) counterparts.\cite{locpaper,VucIroWagTeaGor-PCCP-17} These approaches employ xc energy densities as interpolation input quantities, arising from both the weak and the strong coupling limit. It is well known that the xc energy densities are not uniquely defined and thus one has to be specific on their definition or {\it gauge}. A gauge often considered and discussed in DFT is the one of the electrostatic potential of the exchange-correlation hole and that gauge has been used in the local interpolation schemes so far.\cite{locpaper,ZhoBahErn-JCP-15,BahZhoErn-JCP-16,VucIroWagTeaGor-PCCP-17} However, as said, this gauge is not unique and it is one of the purposes of this work to analyse other possibilities, with particular focus on the one that arises from the recent work of Levy and Zahariev (LZ).\cite{LevZah-PRL-14,LevZah-MP-16,ZahLev-PCA-16}

In the LZ approach the usual Hartree-exchange-correalation (Hxc) potential (that vanishes at infinity for finite systems) is augmented by a density dependent shift, which we call here the LZ constant. The appealing feature of this approach is that the total ground state energy is equal to the sum of the corresponding KS occupied orbital energies, allowing one to model the xc potential without using line integrals to recover the associated energy. In addition to that, it has been shown that upon any isoelectronic changes in the density, the LZ potential changes less on average than the corresponding usual Hxc potential.\cite{LevZah-PRL-14} Furthermore, while it is well known that the latter potential exhibits a discontinuity with the onset of fractional particle number,\cite{PerParLevBal-PRL-82} the former does not exhibit that feature.\cite{LevZah-PRL-14} It has also been shown that the shift appearing in the LZ potential arises very naturally in the strong coupling limit of KS DFT, with a physically transparent meaning.\cite{VucWagMirGor-JCTC-15} 

In this work we analyse some properties of the LZ potential that can be useful to build approximations that are not biased towards weak correlation, such as its strong coupling limit, scaling constraints on the LZ shift at weak and strong correlation, its relation to the response potential, and the corresponding energy density at different coupling strenghts. We also show that the LZ constant can be used to analyse the classical uniform electron gas. We then consider the virial gauge and we show explicitly that the well-known Levy-Perdew virial relation for the exchange potential\cite{LevPer-PRA-85} also holds for the xc potential in the strong coupling limit, generalising to any number of electrons $N$ and any geometry the original proof of Seidl\cite{Sei-PRA-99} for $N=2$ in the case of spherically symmetric densities.
Finally, we analyze the effective charge associated to the exchange-correlation hole at different coupling strengths.\cite{XavGuz-PRL-11,GidLat-JCP-12}

Hartree atomic units are used throughout the paper. 

\section{Theoretical Background}
\label{sec_back}

In KS DFT, the ground state energy and density of a system with an external potential $v$ are given by:
\begin{equation}\label{eq:gs}
E_{\rm GS}= \min_\rho \left\lbrace T_s[\rho]+\int v(\rv) \rho (\rv) \mathrm{d}\rv +E_{\rm Hxc}[\rho]\right\rbrace,
%E_{\rm GS}= \min_\Phi \left\lbrace \langle\Phi|\hat{T}_s[\rho]+\int v(\rv) \rho (\rv) \mathrm{d}\rv +E_{\rm Hxc}[\rho]\right\rbrace,
\end{equation}
where $T_s[\rho]$ is the KS noninteracting kinetic energy functional,\cite{KohSha-PR-65,Lev-PNAS-79}
\begin{equation}\label{eq:ts}
T_s[\rho]= \min_{\Psi \to \rho}\langle \Psi  | \hat{T}| \Psi\rangle,
\end{equation}
and $E_{\rm Hxc}[\rho]$ represents the sum of the Hartree $U[\rho]$ and the xc functional $E_{xc}[\rho]$. For any practical purposes $E_{xc}[\rho]$ must be approximated. 
In KS DFT the minimization in Eq.~\eqref{eq:gs} is carried out by varying the orbitals of the single Slater determinant that usually satisfies Eq.~\eqref{eq:ts},  leading to the single--particle KS equations,
\begin{align}
\Big[&-\frac{1}{2}\nabla^2+v(\rv)+v_{\rm Hxc}([\rho];\rv)\Big]\phi_i(\rv)=\epsilon_i\phi_i(\rv),\\ \nonumber
&\rho(\rv)=\sum_{i=1}^{\rm occ}|\phi_i(\rv)|^2, 
\end{align}
where $v_{\rm Hxc}([\rho];\rv)$ is the functional derivative of $E_{\rm Hxc}[\rho]$ with respect to the density, supplemented by the condition $v_{\rm Hxc}([\rho];|\rv|\to\infty)=0$.

The density-fixed adiabatic connection (AC) formalism\cite{LanPer-SSC-75,GunLun-PRB-76} provides an exact expression for $E_{\rm xc}[\rho]$, by introducing the functional $F_\lambda[\rho]$, with $\lambda$ a real and positive coupling constant,\cite{Lev-PNAS-79}
\begin{equation}
	F_\lambda[\rho]=\min_{\Psi\to\rho}\langle\Psi|\hat{T}+\lambda\,\hat{V}_{ee}|\Psi\rangle.
	\label{eq:Flambda}
\end{equation} 
By denoting $\Psi_{\lambda}[\rho]$ the minimizing wavefunction in Eq.~\eqref{eq:Flambda}, we have
\begin{equation}
E_{xc}[\rho]= \int_0^1 W_\lambda[\rho]\mathrm{d}\lambda,
		\label{eq:10.ac_xc}
\end{equation}
where $W_\lambda[\rho]$ is the global AC integrand,
\begin{equation}
W_\lambda[\rho]= \langle \Psi_{\lambda}[\rho]  | \hat{V}_{ee}| \Psi_{\lambda}[\rho]\rangle - U[\rho].
		\label{eq:10.w_lam}
\end{equation}
We can write $W_\lambda[\rho]$ in terms of the $\lambda$-dependent energy densities, $w_\lambda(\rv)$:
		\begin{align}
W_\lambda[\rho]=\int w_\lambda(\rv) \rho(\rv) \mathrm{d}\rv.
  \label{eq:global}
	\end{align}
The energy density $w_\lambda(\rv)$ is not uniquely defined, as adding to it any function that integrates to zero when multiplied by the density would not change the value of the global quantities. Therefore, for meaningful comparison of energy densities at different $\lambda$, one has to be specific on their definition (or {\it gauge}). One of the most common gauges in DFT is the one of the electrostatic potential of the x/xc hole,\cite{Bec-JCP-05,BecJoh-JCP-07,PerStaTaoScu-PRA-08,MirSeiGor-JCTC-12,locpaper} which has just recently been used for the constructions of xc functionals via local interpolation along the adiabatic connection.\cite{locpaper}. The energy densities within this gauge, which we denote as $w_\lambda^{\rm hole}(\rv)$, are given by:
\begin{equation}\label{eq:wxc_def}
   w_\lambda^{\rm hole}(\rv) = \frac{1}{2} \int \frac{h_{\rm xc}^{\lambda}(\mathbf{r},\mathbf{r}')}{|\mathbf{r} - \mathbf{r}'|} \,\mathrm{d} \mathbf{r}',
\end{equation}
where $h_{xc}^{\lambda}(\rv,\rv')$ is the the xc hole defined as:
     \begin{equation}
h_{xc}^{\lambda}(\rv,\rv')=\frac{P_2^\lambda(\rv,\rv')}{\rho(\rv)}-\rho(\rv'),
\label{eq_hxclambda}
\end{equation}
with  $P_2^\lambda(\rv,\rv')$ the pair--density of $\Psi_\lambda$,
\begin{equation}\label{eq:pair}\begin{aligned}
P_{2}^{\lambda}&(\mathbf{r},\mathbf{r}') = N(N-1) \times \\
& \sum_{\sigma_{1}\ldots\sigma_{N}} \int |\Psi_{\lambda}(\mathbf{r}\sigma_{1},\ldots,\mathbf{r}_{N}\sigma_{N})|^{2} \, \mathrm{d} \mathbf{r}_{3} \ldots \mathrm{d} \mathbf{r}_{N}.
\end{aligned}\end{equation}

%where ${\bar w}_{\rm xc}(\rv)$ is the coupling constant averaged xc energy density in the gauge of the electrostatic potential of the xc hole,\cite{Bec-JCP-05,BecJoh-JCP-07,PerStaTaoScu-PRA-08,MirSeiGor-JCTC-12,locpaper}
%\begin{align}
%\bar w_{\rm xc}(\rv) = \frac{1}{2}\int_0^1 \mathrm{d} \lambda  \int \frac{h_{\rm xc}^{\lambda}(\mathbf{r},\mathbf{r}')}{|\mathbf{r} - \mathbf{r}'|} \, \mathrm{d} \mathbf{r}',
%  \label{eq:wbar}
%	\end{align}

%\begin{align}
% E_{\rm xc}[\rho]=\int \bar w_{xc}(\rv) \rho(\rv) \mathrm{d}\rv.
%  \label{eq:exc}
%	\end{align}

\section{The augmented potential of Levy and Zahariev}
\label{sec_levzah} 

As mentioned, there are several other definitions of the DFT energy densities proposed in the literature.\cite{BurCruLam-JCP-98,CruLamBur-JPCA-98,LevZah-PRL-14} A very recent definition for the Hartree-exchange-correlation energy density comes from the augmented potential, proposed by Levy and Zahariev,\cite{LevZah-PRL-14} which shifts $v_{\rm Hxc}([\rho];\rv)$ by the constant $C[\rho]$,
\begin{align}
v^{\rm LZ}_{\rm Hxc}([\rho];\rv)=v_{\rm Hxc}([\rho];\rv) + C[\rho],
  \label{eq:augpot}
	\end{align}
in such a way that the ground state energy is equal to the sum of the corresponding KS occupied orbital energies $\epsilon_i^{\rm LZ}$, $E_{\rm GS}=\sum_i^N \epsilon_i^{\rm LZ}$. The constant $\lim_{|\mathbf{r}| \to \infty}v^{\rm LZ}_{\rm Hxc}([\rho];\mathbf{r})=C[\rho]$ must then be equal to
\begin{align}
C[\rho]=\frac{E_{\rm Hxc}[\rho]-\int v_{\rm Hxc}([\rho];\rv)\rho(\rv) \mathrm{d}\rv}{\int \rho(\rv)\mathrm{d}\rv}.
  \label{eq:const}
	\end{align}
Of course Eq.~\eqref{eq:const} is not useful if one already has an approximation for $E_{\rm Hxc}[\rho]$; the point of the LZ approach is that one should try to directly build approximations for $v^{\rm LZ}_{\rm Hxc}([\rho];\rv)$, avoiding the need of line integrals. Studying the exact properties of  $v^{\rm LZ}_{\rm Hxc}([\rho];\rv)$ from Eq.~\eqref{eq:const} can provide guiding principles for the construction of such approximations.
Using Eq.~\eqref{eq:const}, we can partition $C[\rho]$ into the Hartree, exchange and correlation components (we drop from now on the argument $[\rho]$ in the potentials),
\begin{align}
C[\rho]&=\underbrace{\frac{U[\rho]-\int v_{\rm H}(\rv)\rho(\rv) \mathrm{d}\rv}{\int \rho(\rv)\mathrm{d}\rv}}_{C_{\rm H}[\rho]=-U[\rho]/N}+\underbrace{\frac{E_x[\rho]-\int v_{x}(\rv)\rho(\rv) \mathrm{d}\rv}{\int \rho(\rv)\mathrm{d}\rv}}_{C_x[\rho]} \nonumber \\
& +\underbrace{\frac{E_c[\rho]-\int v_c(\rv)\rho(\rv) \mathrm{d}\rv}{\int \rho(\rv)\mathrm{d}\rv}}_{C_c[\rho]},
\label{eq:const_partt}
\end{align}
where $v_{\rm H}(\rv)$ is the Hartree potential.
As shown above, $C[\rho]$ consists of the Hartree and the xc component. The $C_{xc}[\rho]$ component of $C[\rho]$ is then equal to
\begin{align}
C_{xc}[\rho]=\frac{E_{\rm xc}[\rho]-\int v_{\rm xc}(\rv)\rho(\rv) \mathrm{d}\rv}{\int \rho(\rv)\mathrm{d}\rv}.
  \label{eq:const_xc}
\end{align}
Multiplying the xc part of the augmented potential of Eq.~\eqref{eq:augpot} by the density and integrating over all space we obtain the xc energy,
\begin{align}
 E_{\rm xc}[\rho]=\int v^{\rm LZ}_{\rm xc}(\rv) \rho(\rv) \mathrm{d}\rv.
  \label{eq:augpotxc}
	\end{align}
From Eq.~\eqref{eq:augpotxc} we see that $v^{\rm LZ}_{\rm xc}(\rv)$, besides being a functional derivative of $E_{\rm xc}[\rho]$ for isoelectronic changes in $\rho$, it also defines an energy density that will be compared with other definitions in Sec.~\ref{sec_edens}. 
%In the next Sec.~\ref{sec_sce} we will focus on the non-trivial LZ shift in the strong coupling limit.

\subsection{Connection between the augmented potential and the response potential}
\label{sec_vresp}
Before moving to the study of the shift in the $\lambda\to\infty$ limit, we also give a simple relation between the LZ constant and the response potential studied by Baerends and coworkers (see, e.g., Refs.~\onlinecite{LeeGriBae-ZPA-95,GriBae-PRA-96,GriMenBae-JCP-16}). We start from the exact equation
\begin{align}
 E_{\rm xc}[\rho]=\frac{1}{2}\int\int \rho(\rv) \rho(\rv') \frac{\bar{g}_{xc}(\rv,\rv')}{|\rv-\rv'|}\mathrm{d}\rv \mathrm{d}\rv',
  \label{eq:ehxc}
	\end{align}
where $\bar{g}_{xc}(\rv', \rv'')$ is the coupling constant averaged (CCA) pair-correlation function, which can be expressed in terms of the CCA exchange-correlation hole,
\begin{align}
\rho(\rv') \bar{g}_{xc}(\rv, \rv')=\int_0^1 h_{xc}^{\lambda}(\rv,\rv') \dd \lambda.
  \label{eq:excG}
	\end{align}
Taking the functional derivative of $E_{\rm xc}[\rho]$ expressed by Eq.~\eqref{eq:ehxc} with respect to the density, we can partition $v_{\rm xc}(\rv)$ as\cite{LeeGriBae-ZPA-95,GriBae-PRA-96,GriMenBae-JCP-16}
\begin{align}
v_{xc}(\rv)= \bar{v}_{\rm resp}(\rv)+2\,\bar{w}_{\rm xc}^{\rm hole}(\rv),
  \label{eq:part}
	\end{align}
where $\bar{w}_{\rm xc}^{\rm hole}(\rv)$ is the coupling constant averaged xc energy density in the gauge of the electrostatic potential of the xc hole,\cite{Bec-JCP-05,BecJoh-JCP-07,PerStaTaoScu-PRA-08,MirSeiGor-JCTC-12,locpaper}
\begin{align}
\bar{w}_{\rm xc}^{\rm hole}(\rv) = \frac{1}{2}\int_0^1 \mathrm{d} \lambda  \int \frac{h_{\rm xc}^{\lambda}(\mathbf{r},\mathbf{r}')}{|\mathbf{r} - \mathbf{r}'|} \, \mathrm{d} \mathbf{r}',
  \label{eq:wbar}
	\end{align}
and ${\bar v} _{\rm resp}(\rv)$ is the response potential that includes the kinetic contribution via the coupling-constant average\cite{LeeGriBae-ZPA-95,GriBae-PRA-96,GriMenBae-JCP-16}
\begin{align}
{\bar v}_{\rm resp}(\rv)=\frac{1}{2} \iint  \frac{\rho(\rv')\rho(\rv'')}{| \rv' - \rv'' |} \frac{\delta \bar{g}_{xc}(\rv', \rv'')}{\delta\rho(\textbf{r})} \mathrm{d}\rv' \mathrm{d}\rv''.
\label{eq:vresp}
\end{align}
The potential ${\bar v}_{\rm resp}(\rv)$ can be interpreted as a measure of the sensitivity of the pair-correlation function to density variations.\cite{LeeGriBae-ZPA-95,GriBae-PRA-96} The response potential is a part of the xc potential that does not contribute to the xc energy, since the xc functional can be written only in terms of the second term appearing on the right-hand side of Eq.~\eqref{eq:part},
\begin{align}
 E_{\rm xc}[\rho]=\int \bar w_{xc}^{\rm hole}(\rv) \rho(\rv) \mathrm{d}\rv.
  \label{eq:exc}
	\end{align}
Plugging Eq.~\eqref{eq:part} into Eq.~\eqref{eq:augpot} and using the definitions of Eq.~\eqref{eq:const_partt} we obtain for the xc part
\begin{align}
v^{\rm LZ}_{\rm xc}(\rv)={\bar v} _{\rm resp}(\rv)+2{\bar w}_{xc}^{\rm hole}(\rv)+ C_{xc}[\rho].
  \label{eq:augpot2}
	\end{align}
By multiplying both sides of this equation by the density $\rho(\rv)$ and integrating over all space, we can find a relationship between the non trivial part $C_{xc}[\rho]$ of the constant $C[\rho]$ and the expectation value $\bar{V}_{\rm resp}[\rho]$ of the response potential,\cite{GriMenBae-JCP-16} defined as 
\begin{align}
\bar{V}_{\rm resp}[\rho]=\int {\bar v}_{\rm resp}(\rv) \rho(\rv) \mathrm{d}\rv. 
\end{align}
This relationship reads as
\begin{align}
C_{xc}[\rho]=-\frac{E_{xc}[\rho]+\bar{V}_{\rm resp}[\rho]}{N},
  \label{eq:CN}
	\end{align}
and shows that one could approximate the constant being guided by the sum of average properties of the response potential and of the xc functional.\cite{LeeGriBae-ZPA-95,GriBae-PRA-96,GriMenBae-JCP-16,KohPolSta-PCCP-16}

\section{The constant $C[\rho]$ in the strong coupling limit of density functional theory}
\label{sec_sce}

The strictly-correlated electrons (SCE) functional is the natural counterpart of the non--interacting KS kinetic energy functional given in Eq.~\eqref{eq:ts}.\cite{Sei-PRA-99,SeiGorSav-PRA-07,GorSeiVig-PRL-09,SeiVucGor-MP-16} It is defined by the following constrained minimization:\cite{SeiGorSav-PRA-07,GorSeiVig-PRL-09,SeiPerLev-PRA-99} 
	\begin{align}
		V_{ee}^{\rm SCE}[\rho]= \inf_{\Psi \to \rho}\langle \Psi  | \hat{V}_{ee}| \Psi\rangle,
		\label{eq:vee_sce}
	\end{align} 
and gives the $\lambda\to\infty$ limit of the density-fixed adiabatic connection\cite{SeiGorSav-PRA-07,SeiPerLev-PRA-99} of Eq.~\eqref{eq:10.w_lam}.
A candidate for the minimizing $|\Psi_\infty[\rho]|^2$  is a distribution parametrized by the so-called {\em co-motion functions} $\fv_i(\rv)$,\cite{SeiGorSav-PRA-07, MirSeiGor-JCTC-12} with a simple physical meaning: if a reference electron is found at $\rv$, then $\rv_i=\fv_{i}(\rv)$ determines the position of all the other electrons in the system.\cite{SeiGorSav-PRA-07} In terms of the co-motion functions, the SCE functional $V_{ee}^{\rm SCE}[\rho]$ is given by\cite{MirSeiGor-JCTC-12}
\begin{align}
	V_{ee}^{\rm SCE}[\rho]=\inf_{\{\fv_n\}:\rho} \int \frac{\rho(\rv)}{2}\sum_{i=2}^{N}\frac{1}{\left | \rv-\fv_i(\rv) \right|}\mathrm{d}\rv,
\label{eq:vee_comotion}
	\end{align}
with the $\fv_i(\rv)$ satisfying group properties
\begin{align}
& \fv_1(\rv)\equiv \rv, \nonumber\\
&  \fv_2(\rv)\equiv \fv(\rv), \nonumber\\
& \fv_3(\rv)=      \fv(\fv(\rv)), \nonumber\\
&\vdots& \label{eq:groupprop}\\
& \fv_N(\rv)=      \underbrace{\fv(\fv(\dotso\fv(\rv)\dotso))}_\text{$N\!-\!1$ times}   ,\nonumber\\
& \underbrace{\fv(\fv(\dotso\fv(\rv)\dotso))}_\text{$N$ times} = \rv,\nonumber
\end{align}
and the constraint ``$\{\fv_n\}:\rho$'' meaning that the co-motion functions satisfy the equation
\begin{align}
\rho \big (\fv(\mathbf{r}) \big)\,J(\fv(\mathbf{r}))  = \rho (\mathbf{r}),
  \label{eq:difcmf}
	\end{align}
where $J(\fv_i(\rv))$ is the Jacobian of the transformation $\rv\to \fv_i(\rv)$ (see Ref.~\citenum{SeiDiMGerNenGieGor-PRA-17} for a recent review). The co-motion functions ansatz has been proven\cite{ColDiM-INC-13} to be exact, but it might happen that it yields only an infimum and not a minimum.\cite{ColDiM-INC-13,SeiDiMGerNenGieGor-PRA-17}
Even though the SCE functional has an ultra nonlocal character, we can easily compute its functional derivative, the SCE potential $v_{\rm SCE}(\rv)$, using the following exact relation:\cite{MalGor-PRL-12,ButDepGor-PRA-12}
	\begin{align}
		\nabla v_{\rm SCE}(\rv)=-\sum_{i=2}^{N}\frac{\rv-\fv_i(\rv)}{\left | \rv-\fv_i(\rv) \right|^3},
		\label{eq:pot_sce}
	\end{align}
which defines the potential up to a constant.
As usual for systems with a fixed number of particles, this constant is fixed by the condition that $v_{\rm SCE}(\rv)$ vanishes when $|\rv|\to\infty$.\cite{MalGor-PRL-12,MalMirCreReiGor-PRB-13,MenMalGor-PRB-14}

In the strong coupling limit, the constant $C[\rho]$ arises very naturally from the SCE functional\cite{VucWagMirGor-JCTC-15}  using the dual Kantorovich formulation\cite{ButDepGor-PRA-12} that provides an alternative expression for  $V_{ee}^{\rm SCE}[\rho]$,
	\begin{align}
& \label{eq:vee_kant} V_{ee}^{\rm SCE}[\rho]=  \\ \nonumber
&\max_{u} \left \{  \int u(\rv)\rho (\rv)\dd \rv:\sum_{i=1}^{N}u(\rv_i)\leqslant \sum_{i=1}^{N}\sum_{j>i}^{N}\frac{1}{ |  \rv_{i}-\rv_{j} |}\right \}
	\end{align}
The Kantorovich potential $u(\rv)$, which achieves the maximum in Eq.~\eqref{eq:vee_kant}, differs from the SCE potential only by a constant and that constant is exactly the one appearing in Eq.~\eqref{eq:augpot} in the strong coupling limit:\cite{VucWagMirGor-JCTC-15} 
	\begin{align}
C_{\rm{SCE}}[\rho]=u(\rv)-v_{\rm SCE}(\rv).
		\label{eq:csceU}
		\end{align}
To understand the meaning of this constant, we invoke again the $\lambda$-dependent Hohenberg-Kohn functional $F_\lambda[\rho]$ given in Eq.~\eqref{eq:Flambda}. If the density $\rho$ is $v$-representable for all $\lambda$, the Lagrange multiplier associated with the constraint $\Psi\to\rho$ yields a one-body potential $\hat{V}_\lambda[\rho]$, defining a series of $\lambda$-dependent hamiltonians $\hat{H}_{\lambda} = \hat{T} + \lambda\hat{V}_{ee} + \hat{V}_\lambda[\rho]$, whose ground-state wavefunctions $\Psi_\lambda$ have all the same density $\rho$.
In the limit $\lambda\to\infty$, $\hat{H}_{\lambda}$ becomes classical,\cite{SeiGorSav-PRA-07,GorVigSei-JCTC-09} 
\begin{align}
\hat{H}_{\lambda \to \infty}=\lambda(\hat{V}_{ee}-\hat{V}_{\rm SCE}).
		\label{eq:HSCE}
	\end{align} 
This Hamiltonian defines a classical electrostatic problem with a degenerate minimum on a 3-dimensional manyfold of the full 3$N$-dimensional configuration space. The manyfold is parametrized by the co-motion functions.\cite{SeiGorSav-PRA-07,GorVigSei-JCTC-09} As the total energy of this system is given by $N\,C_{\rm{SCE}}[\rho]$, we can write $C_{\rm{SCE}}[\rho]$ as\cite{VucWagMirGor-JCTC-15}
\begin{equation}
	C_{\rm{SCE}}[\rho]=\frac{1}{N}\lim_{\lambda\to\infty}\frac{\langle\Psi_\lambda[\rho]|\hat{H}_\lambda|\Psi_\lambda[\rho]\rangle}{\lambda}.
	\label{eq:CSCE}
\end{equation}
Taking the limit in this expression,\cite{GorVigSei-JCTC-09}  we can see that $C_{\rm{SCE}}[\rho]$ represents the electrostatic energy per electron (within the standard gauge in which all the potentials appearing in the hamiltonian are set to zero at infinity):\cite{VucWagMirGor-JCTC-15}
\begin{align}
C_{\rm{SCE}}[\rho] & = \min_{\rv_1,...\rv_N}\frac{\sum_{j>i}^{N}\frac{1}{\left | \rv_i-\rv_j \right |}-\sum_{i=1}^{N}v_{\rm SCE} (\rv_i)}{N} \nonumber \\
& = \frac{\sum_{j>i}^{N}\frac{1}{\left | \fv_i(\rv)-\fv_j(\rv) \right |}-\sum_{i=1}^{N}v_{\rm SCE} (\fv_i(\rv))}{N},
	\label{eq:CSCE_epot}
\end{align}
where the minimum is degenerate in $\rv$, so that one can obtain the constant by choosing any value of $\rv$, for example by putting one of the electrons at infinity.\cite{VucWagMirGor-JCTC-15} In the case in which the SCE state only provides an infimum, the same results can be applied to the support of the minimizing distribution.\cite{SeiDiMGerNenGieGor-PRA-17}
This electrostatic meaning of the LZ constant in the SCE limit could be used to build approximations.

\subsection{Scaling of the constant $C[\rho]$ in the weak and strong coupling limit}
\label{sec_scal}
In the SCE limit, Eq.~\eqref{eq:const} becomes\cite{VucWagMirGor-JCTC-15}
\begin{align}
C_{\rm{SCE}}[\rho]=V_{\rm ee}^{\rm SCE}[\rho]-\frac{\int v_{\rm SCE}(\rv)\rho(\rv) \mathrm{d}\rv}{\int \rho(\rv) \mathrm{d}\rv}
  \label{eq:constSCE}
	\end{align}
Defining $\rho_\gamma(\rv)=\gamma^3\rho(\gamma\,\rv)$, with $\gamma>0$,\cite{LevPer-PRA-85,LevPer-PRB-93} we can determine a scaling relation for the $C_{\rm{SCE}} [\rho]$ constant. Knowing how this object scales under uniform coordinate scaling is an important exact constraint for approximating this quantity. The SCE functional and potential satisfy the scaling relations\cite{GorSei-PCCP-10}
\begin{align}
V_{ee}^{\rm SCE}[\rho_\gamma]=\gamma V_{ee}^{\rm SCE}[\rho]
  \label{eq:scalingVee}
	\end{align}
and
\begin{align}
v_{\rm SCE}([\rho_\gamma],\rv)=\gamma v_{\rm SCE}([\rho],\gamma \rv).
  \label{eq:scalingvsce}
	\end{align}
The $\gamma$-scaled $C_{\rm{SCE}}[\rho]$ reads as
\begin{align}
C_{\rm{SCE}}[\rho_\gamma]=V_{\rm ee}^{\rm SCE}[\rho_\gamma]-\frac{\int v_{\rm SCE}([\rho_\gamma],\rv)\rho_\gamma(\rv) \mathrm{d}\rv}{\int \rho_\gamma(\rv). \mathrm{d}\rv}
  \label{eq:scalCsce1}
	\end{align}
Plugging Eqs.~\eqref{eq:scalingVee} and \eqref{eq:scalingvsce} into Eq.~\eqref{eq:scalCsce1}, we obtain the scaling relation for $C_{\rm{SCE}}[\rho]$
\begin{align}
C_{\rm{SCE}}[\rho_\gamma]=\gamma V_{\rm ee}^{\rm SCE}[\rho]-\frac{\gamma\int v_{\rm SCE}(\gamma \rv)\rho_\gamma(\rv) \mathrm{d}\rv}{\int \rho_\gamma(\rv) \mathrm{d}\rv}
  \label{eq:scalCsce2}
	\end{align}
From the previous equation it follows that $C_{\rm{SCE}}[\rho_\gamma]$ obeys the following scaling relation:
\begin{align}
C_{\rm{SCE}}[\rho_\gamma]=\gamma C_{\rm{SCE}}[\rho]
  \label{eq:scalCscef}
	\end{align}
Combining Eqs.~\eqref{eq:csceU}, \eqref{eq:scalingvsce} and~\eqref{eq:scalCscef} we obtain the following scaling relation for the Kantorovich potential, i.e. the augmented Hxc potential in the SCE limit:
\begin{align}
u([\rho_\gamma],\rv)=\gamma u([\rho],\gamma \rv).
  \label{eq:scalCsceU}
	\end{align}
%In Figure~\ref{fig_c_he} we show how $C_{\rm{SCE}}[\rho]/Z$ varies with the nuclear charge $Z$ in the case of the helium isoelectronic series, whose density nearly satisfies the uniform coordinate scaling at large $Z$ with $\gamma \approx Z$, $\rho_Z(\rv) \approx Z^3 \rho(Z \rv)$. We see that when $Z$ gets larger $C_{\rm{SCE}}[\rho]/Z$ approaches a constant, as predicted by the scaling relation. 

%%%%%%%%%%
%\begin{figure}
%\includegraphics[width=6cm]{c_z_he.pdf}
%\caption{The constant $C_{\rm SCE}[\rho]/Z$ in the strong coupling limit versus the nuclear charge $Z$, for the helium isoelectronic series. The densities of the given series have been obtained at the FCI/aug-cc-pCVTZ level of theory}
%\label{fig_c_he}
%\end{figure}
%%%%%%%%%
\begin{table}
\centering
\begin{tabular}{@{}llll@{}}
\toprule
Atom & $C_x[\rho]$                & $C_{xc}[\rho]$ & $C_{xc}^{\rm SCE} [\rho]$ \\ \midrule
\hline \hline
He   & 0.5123 & 0.4765         & 0.5046                    \\
Be   & 0.3648                     & 0.4151         & 0.4797                    \\
Ne   &  0.776                     & 0.681           & 0.894                     \\ \bottomrule
\end{tabular}
\caption{The exchange and correlation components of the $C[\rho]$ constants for He, Be and Ne compared with $C_{xc}^{\rm SCE} [\rho]$}
\label{tab_atoms}
\end{table}	

\subsection{$C[\rho]$ at different coupling strengths for small atoms}
\label{sec_atoms}

Here we study how the xc part of the LZ constant, $C[\rho]+U[\rho]/N$, varies with coupling strength  $\lambda$, by comparing its value at $\lambda=0$, corresponding to the exchange-only $C_x[\rho]$, at full coupling strength $\lambda=1$, corresponding to $C_{xc}[\rho]$ for the physical hamiltonian, and at $\lambda=\infty$, corresponding to $C_{xc}^{\rm SCE}[\rho]=C_{\rm SCE}[\rho]+U[\rho]/N$.

In Table~\ref{tab_atoms} we compare the results for $C_x[\rho]$, $C_{xc}[\rho]$ and $C_{xc}^{\rm SCE}[\rho]$ for the helium, beryllium and neon atoms. The shifts have been obtained by using always the same highly accurate density\cite{UmrGon-PRA-94,SeiGorSav-PRA-07} to evaluate them at different coupling strengths. We have used accurate exchange-correlation and exchange only potentials and energies from the existing literature,\cite{FilUmrGon-PRA-96,UmrGon-PRA-94} to compute $C_x[\rho]$ and $C_{xc}[\rho]$. To compute $C_{xc}^{\rm SCE}[\rho]$, we have used Eq.~\eqref{eq:constSCE}, obtaining $V_{\rm ee}^{\rm SCE}[\rho]$ and $v_{\rm SCE}(\rv)$ with the conjectured SCE solution for spherically symmetric systems proposed in Ref.~\onlinecite{SeiGorSav-PRA-07}, which gives either exact or very accurate SCE quantities.\cite{SeiDiMGerNenGieGor-PRA-17} 
  
As we can see from Table~\ref{tab_atoms} the trends are not regular: for He and Ne the physical $C_{xc}[\rho]$ is lower than both the exchange and the SCE values, while for Be the LZ constant increases with increasing coupling strength. This feature might be linked to the fact that Be has a smaller gap than He and Ne, but of course we have too little data to really draw this conclusion. Further investigation of this aspect will be the object of future works.  

\subsection{Spheres of uniform density}
\label{sec:parlam}
In a recent work, Lewin and Lieb\cite{LewLie-PRA-15} have shown that in three dimensions the {\em total} (Madelung) energy per electron of the bcc Wigner crystal cannot be identified with an exchange-correlation energy, with important implications for the Lieb-Oxford inequality.\cite{LewLie-PRA-15,Lie-PLA-79,LieOxf-IJQC-81}
In this context, SCE calculations on uniform spherical densities can shed light on the behavior of the classical uniform electron gas in the thermodynamic limit. 
Similarly to Refs.~\onlinecite{RasSeiGor-PRB-11,SeiVucGor-MP-16}, we consider here the following spherically-symmetric density profile
\begin{align}
\rho_a(r)=\frac{k_N(a)}{e^{a(r-1)}+1},
  \label{eq:rho_a_uni}
	\end{align}
where $a$ is a parameter, and $k_N(a)$ is a constant that ensures that $\rho_a(r)$ integrates to $N$. In the $a \to \infty$ limit, this density becomes the one of a uniform sphere with radius $R=1$:
\begin{align}
\rho_{\rm unif}(r) = \begin{cases}
\frac{3N}{4 \pi} & r \leq 1 \\
0 & r>1
\end{cases}
  \label{eq:rho_uni}
	\end{align} 
%%%%%%%%%%
\begin{figure}
\includegraphics[width=9cm]{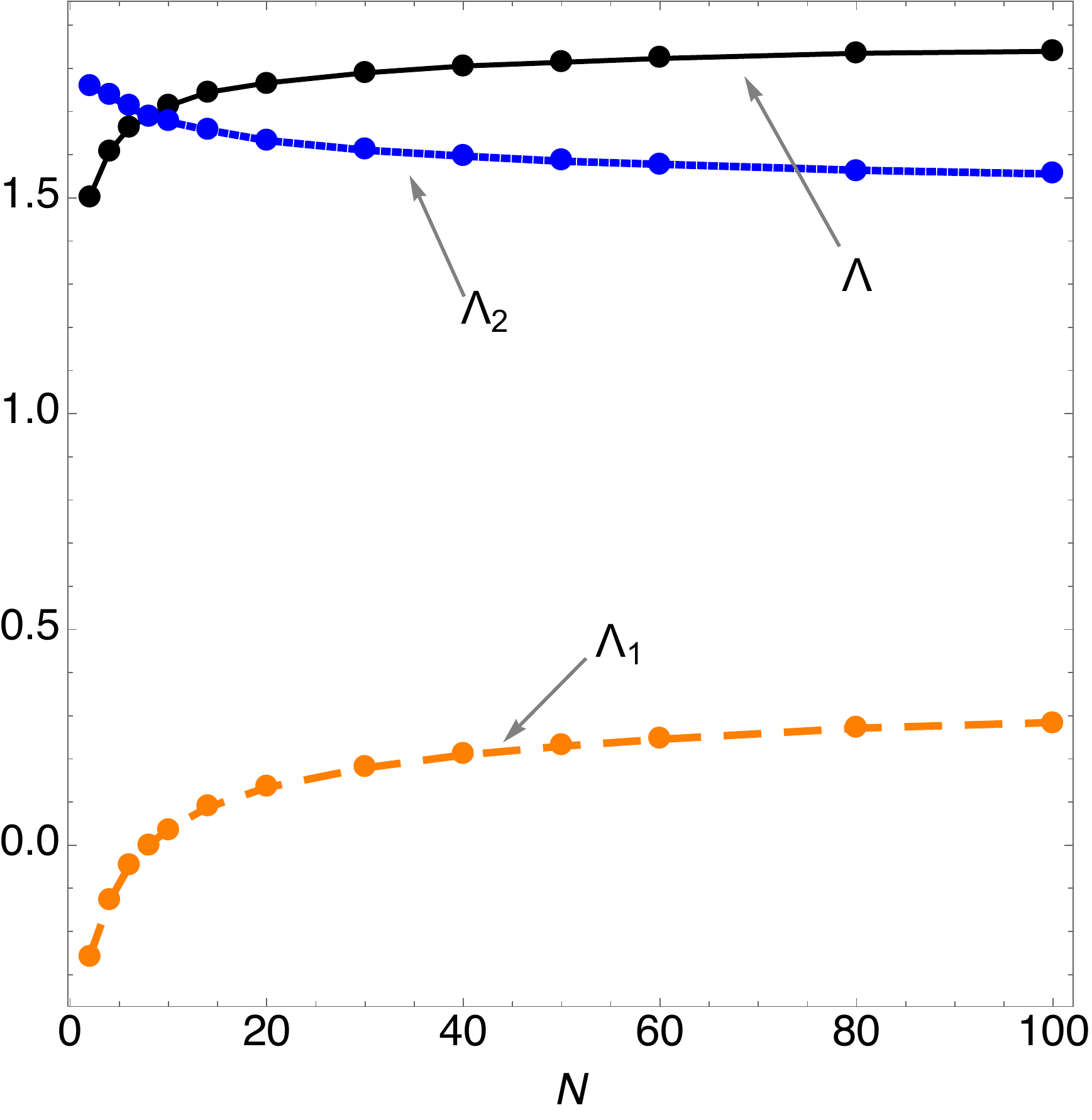}
\caption{$\Lambda_1[\rho]$, $\Lambda_2[\rho]$ and their sum versus $N$ for the quasi-uniform density of Eq.~\eqref{eq:rho_a_uni} with $a=500$.}
\label{fig_lamall1}
\end{figure}
%%%%%%%%%
Using the density profile of Eq.~\eqref{eq:rho_a_uni} with a finite, but very large value of the parameter $a$, we can approach a uniform density, and we can also ensure that the corresponding functional derivative $v_{\rm SCE}([\rho_a];r)$ vanishes at infinity. While more details of these calculations and the implications for the Lieb-Oxford inequality will be reported elsewhere, here the aim is to use the LZ constant to provide more insight into the large-$N$ behavior of the functional\cite{RasSeiGor-PRB-11,SeiVucGor-MP-16}
\begin{align}
	\Lambda[\rho]=\frac{V_{ee}^{\rm SCE}[\rho]-U[\rho]}{E_x^{\rm LDA}[\rho]},
	\label{eq:Lambda}
\end{align}
whose maximum with respect to all possible densities provides the optimal constant appearing in the Lieb-Oxford inequality.\cite{SeiVucGor-MP-16} In Eq.~\eqref{eq:Lambda} $E_x^{\rm LDA}[\rho]=-\frac{3}{4}(\frac{3}{\pi})^{1/3}\int\rho(\rv)^{4/3}\dd\rv$ is the usual local-density exchange functional.

We thus rewrite the SCE functional in terms of its functional derivative (gauged to zero at infinity) and the non-trivial part $C_{\rm xc}^{\rm SCE}[\rho]$ of the LZ shift,
\begin{align}
V_{\rm ee}^{\rm SCE}[\rho]=C_{\rm xc}^{\rm SCE}[\rho] N- U[\rho]+ \int v_{\rm SCE}([\rho];\rv) \rho(\rv) \dd \rv.
  \label{eq:vee_constxc}
	\end{align}
Plugging Eq.~\eqref{eq:vee_constxc} into Eq.~\eqref{eq:Lambda} we obtain:
\begin{align}
\Lambda[\rho]=\underbrace{\frac{C_{\rm xc}^{\rm SCE}[\rho] N}{E_x^{\rm LDA}[\rho]}}_{\Lambda_1[\rho]}+\underbrace{\frac{\int v_{\rm SCE}([\rho];\rv) \rho(\rv) \dd \rv-2U[\rho]}{E_x^{\rm LDA}[\rho]}}_{\Lambda_2[\rho]}.
  \label{eq:Lambda2}
	\end{align}
The functional $\Lambda_2[\rho]$ would go to zero in the thermodynamic limit if the SCE potential approached the potential of a sphere of uniform positive background with the same density $\rho_{\rm unif}$ when $N\to\infty$, $v_{\rm unif}(r)=-\int\frac{\rho_{\rm unif}\,\dd{\bf r}'}{|\rv-\rv'|}$.  The external potential $v_{\rm unif}(r)$ defines the classical jellium model, whose electronic density is in general not uniform and cannot be made uniform in a simple way, even in the thermodynamic limit, due to the long-range nature of the Coulomb interaction.\cite{LewLie-PRA-15}  

The SCE functional reformulates the problem in a different way: the electronic density is now forced to be uniform by the external potential\cite{SeiGorSav-PRA-07,RasSeiGor-PRB-11,MalGor-PRL-12,SeiVucGor-MP-16} $-v_{\rm SCE}([\rho];r)$, which, in general, is not equal to the one created by a uniform positive background (for an in-depth analysis of the difference between jellium and the uniform electron gas, see the recent work of Lewin, Lieb and Seiringer\cite{LewLieSei-arxiv-17}). The LZ shift allows us to isolate and analyze the contribution from the external potential to $\Lambda[\rho]$. 

In Fig.~\ref{fig_lamall1} we show the functional $\Lambda[\rho]$ and its two components of Eq.~\eqref{eq:Lambda2} as a function of $N$ for the densities of Eq.~\eqref{eq:rho_a_uni} with $a=500$: the trivial leading term of  $\int \rho\,v_{\rm SCE} $, which goes like $N^2$, clearly cancels exactly the term $2U[\rho]$ (otherwise $\Lambda_2[\rho]$ would diverge for large $N$). However, the contribution to the next leading order, $\sim N^{4/3}$, which is the crucial one for the Lieb-Oxford bound, is clearly big and does not seem to disappear as $N$ grows.

\begin{figure}
    \includegraphics[width=0.8\linewidth]{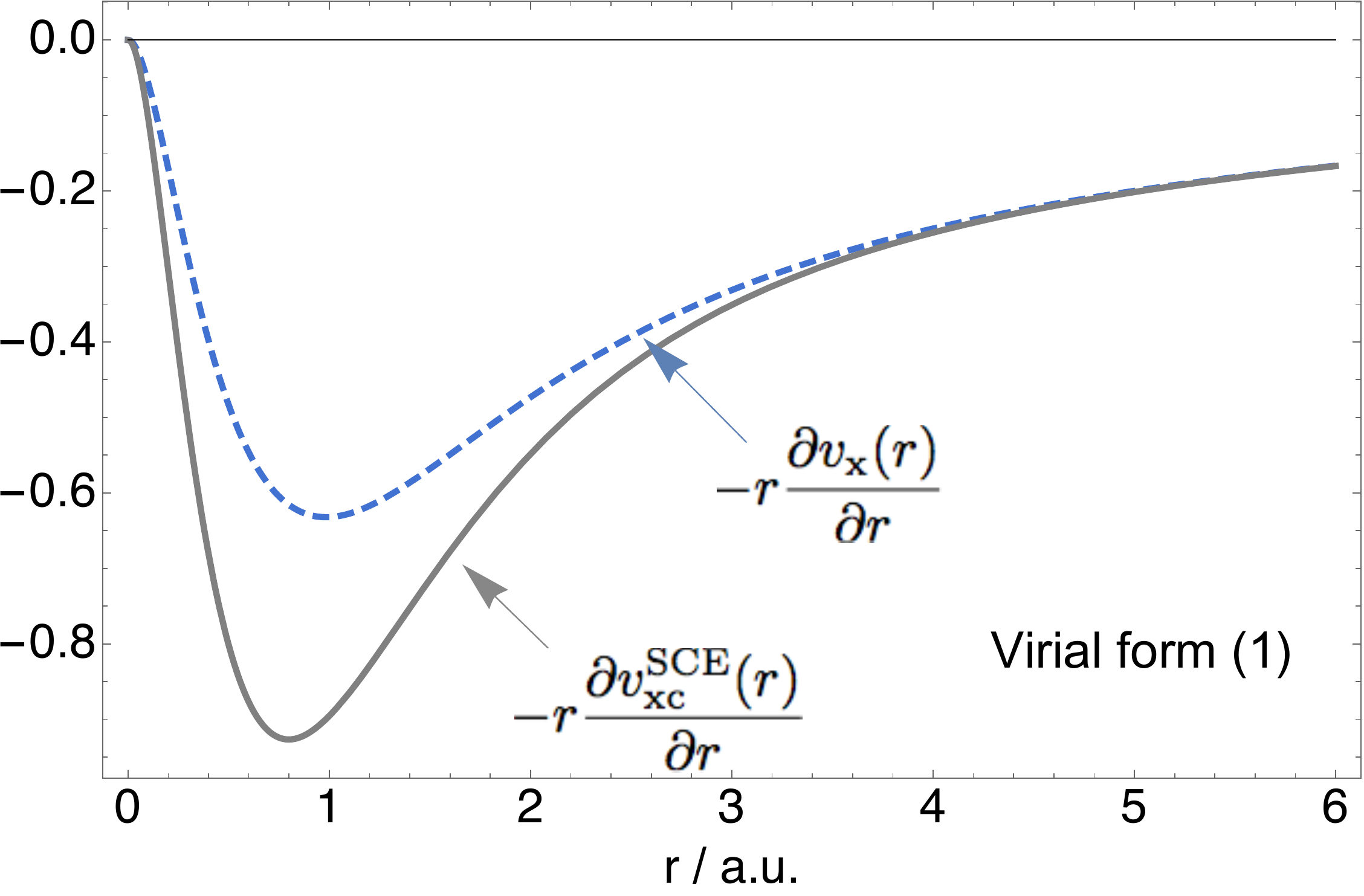}\par
    \includegraphics[width=0.8\linewidth]{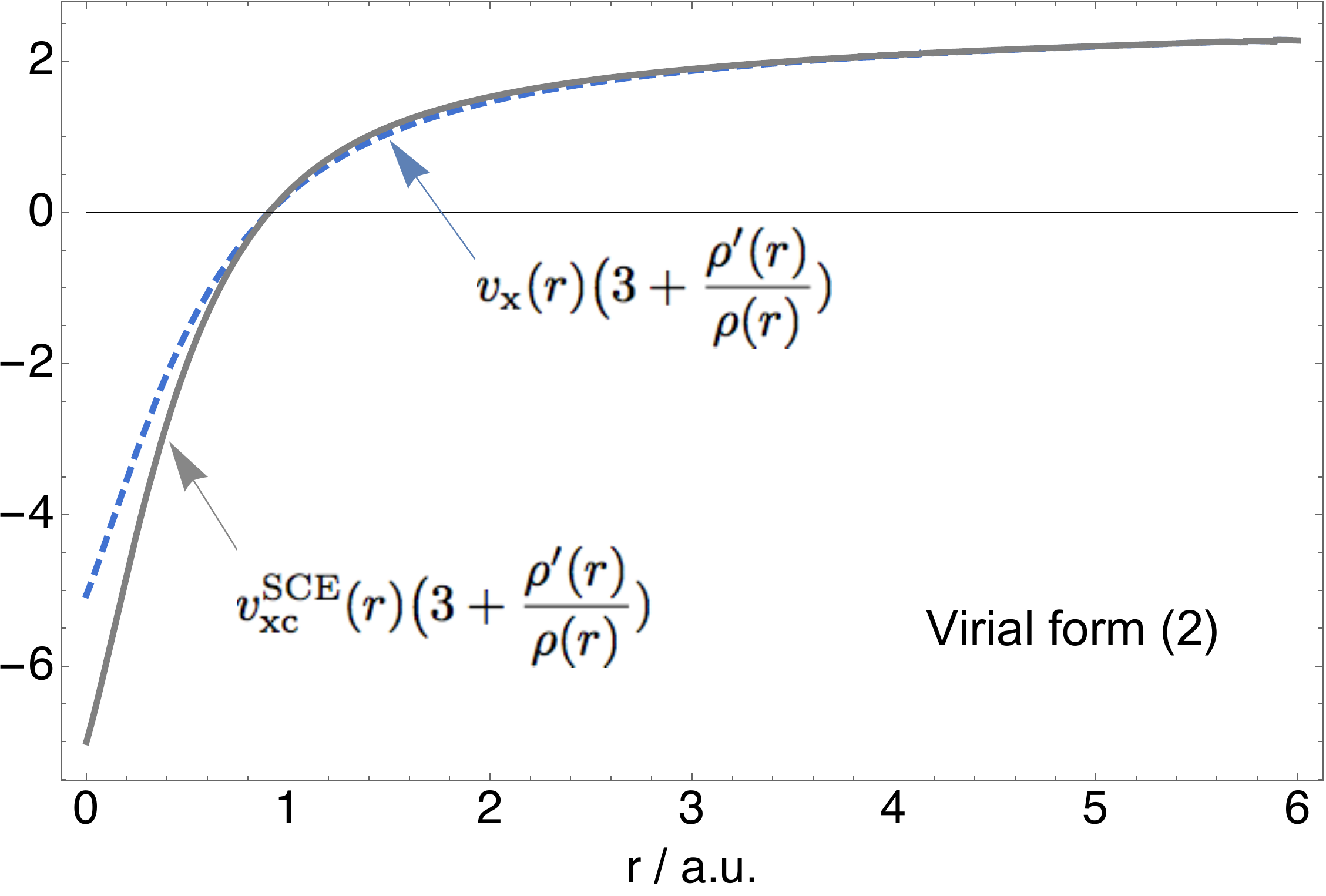}\par
    \includegraphics[width=0.8\linewidth]{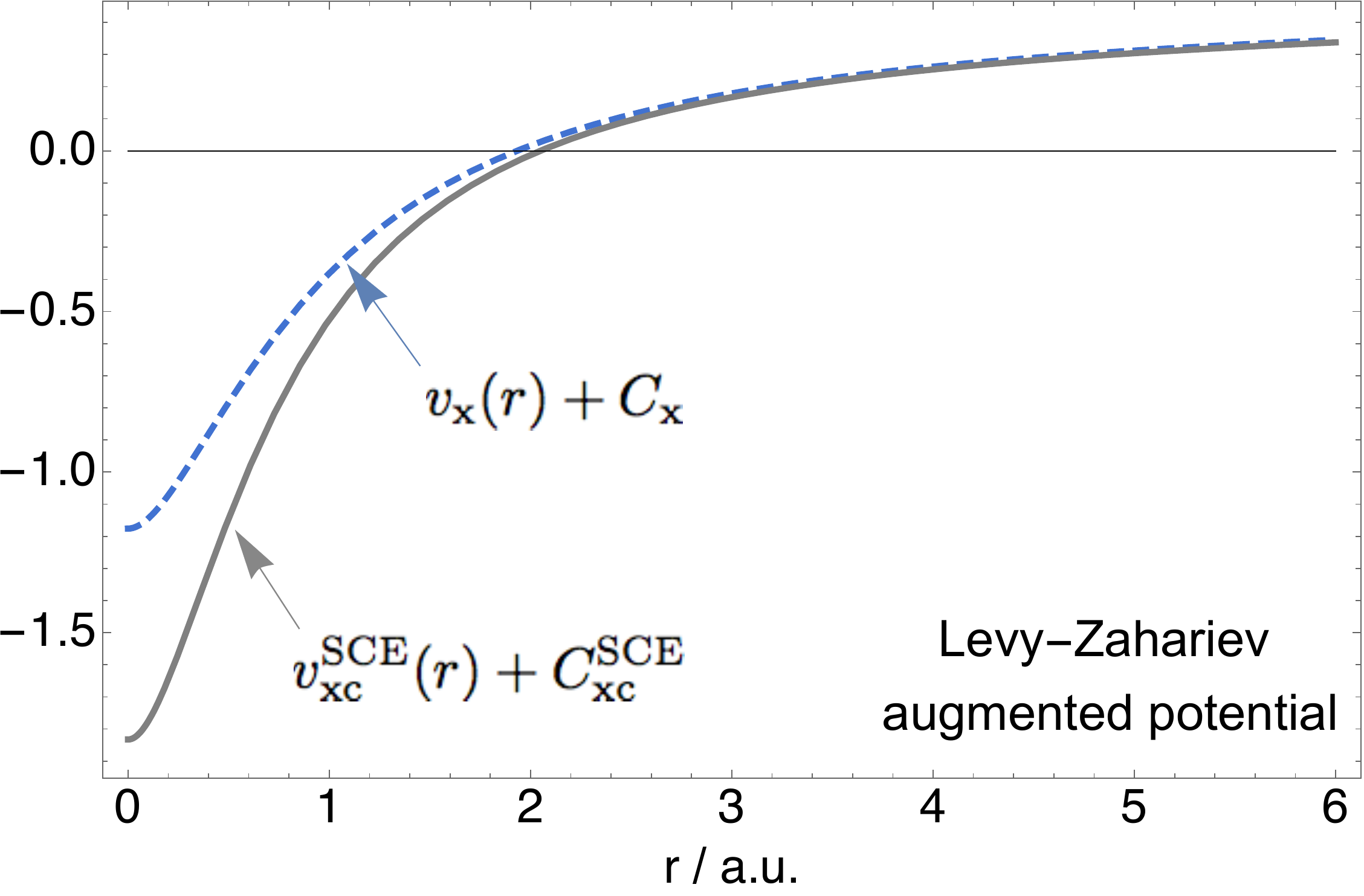}\par
    \includegraphics[width=0.8\linewidth]{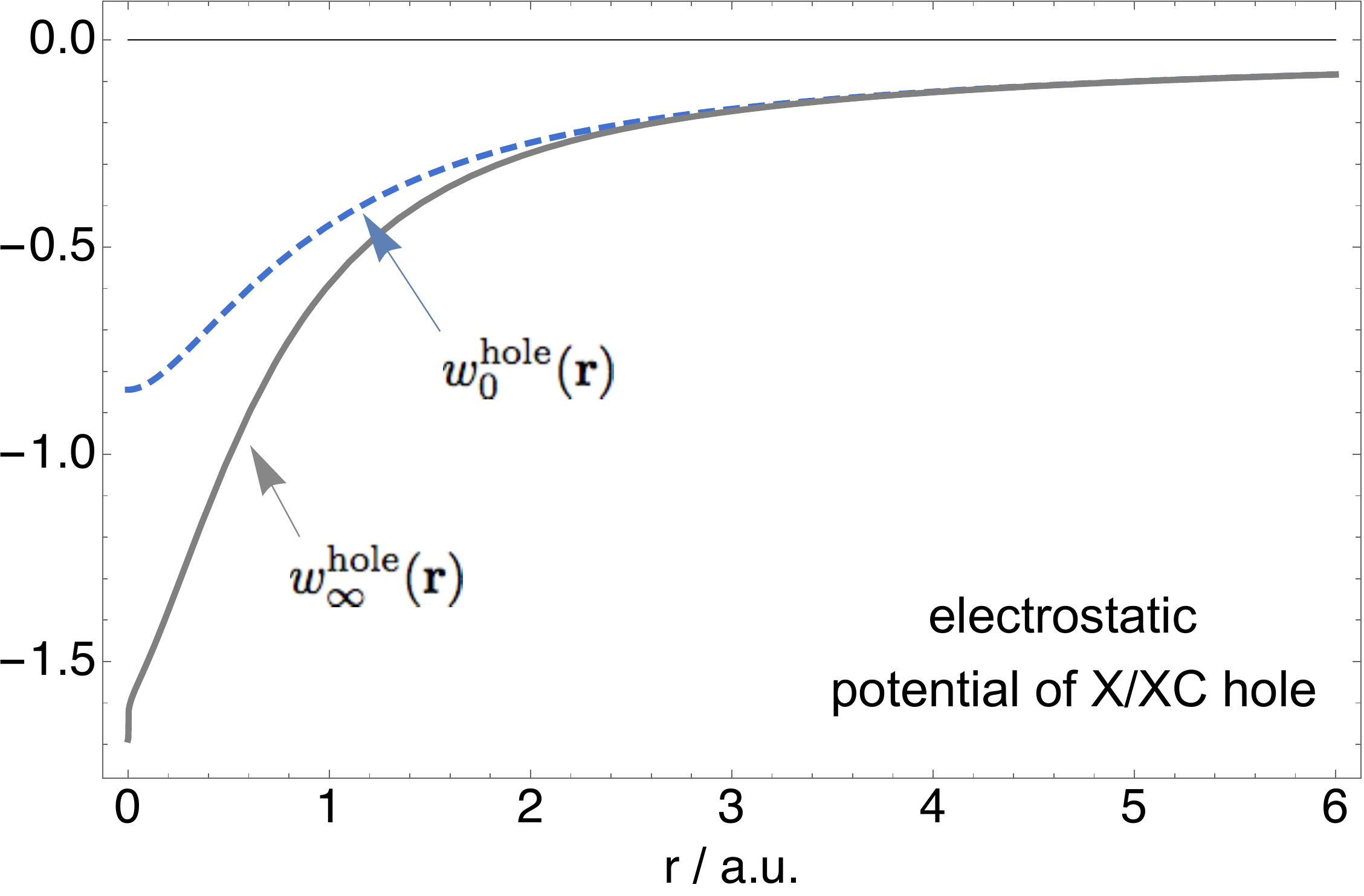}\par
\caption{Weak and strong coupling limit energy densities for the helium atom, within the different energy density gauges of Table~\ref{table_edens}.}
\label{fig_edens_he}
\end{figure}
\begin{figure}
    \includegraphics[width=0.8\linewidth]{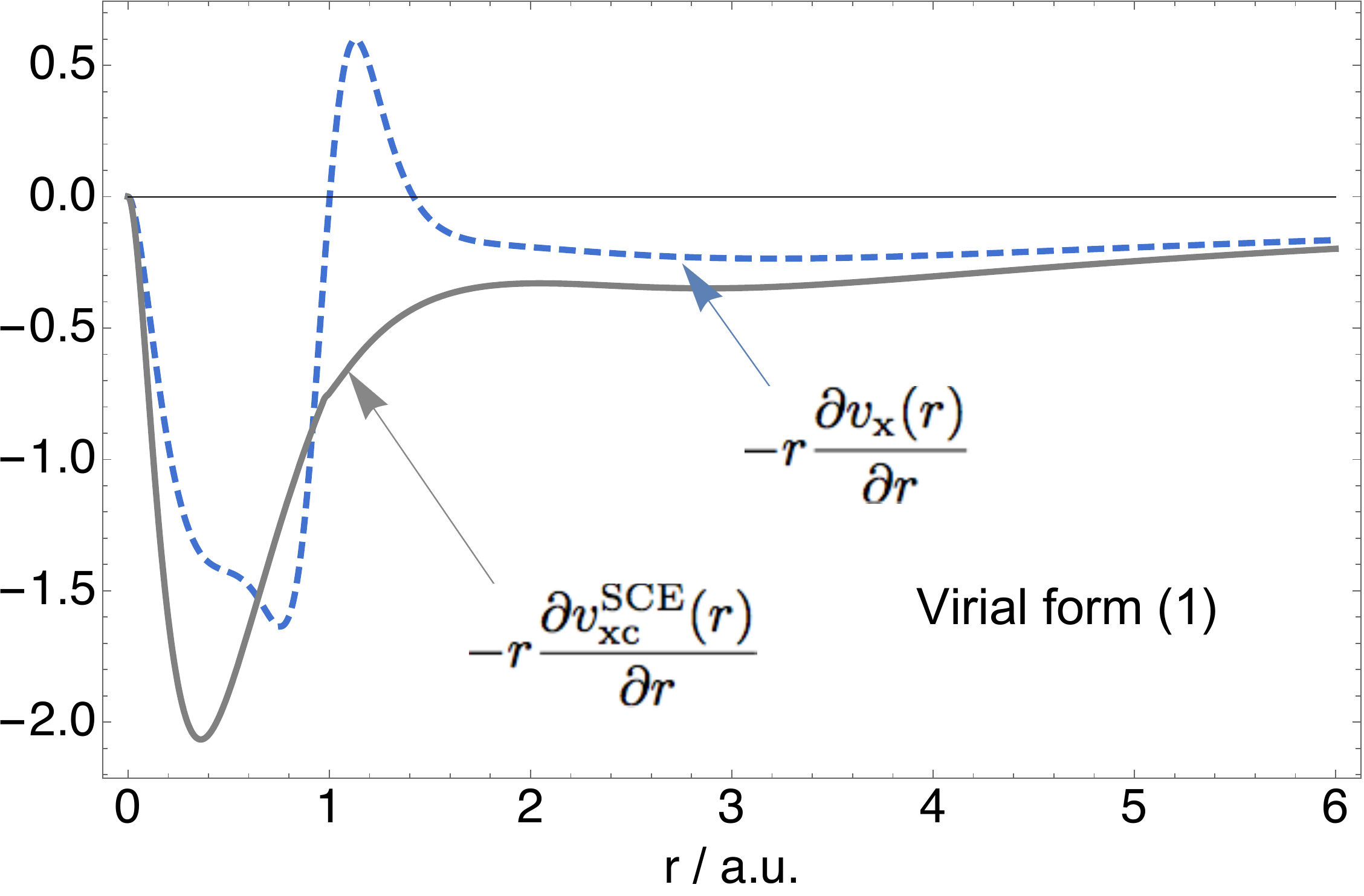}\par
    \includegraphics[width=0.8\linewidth]{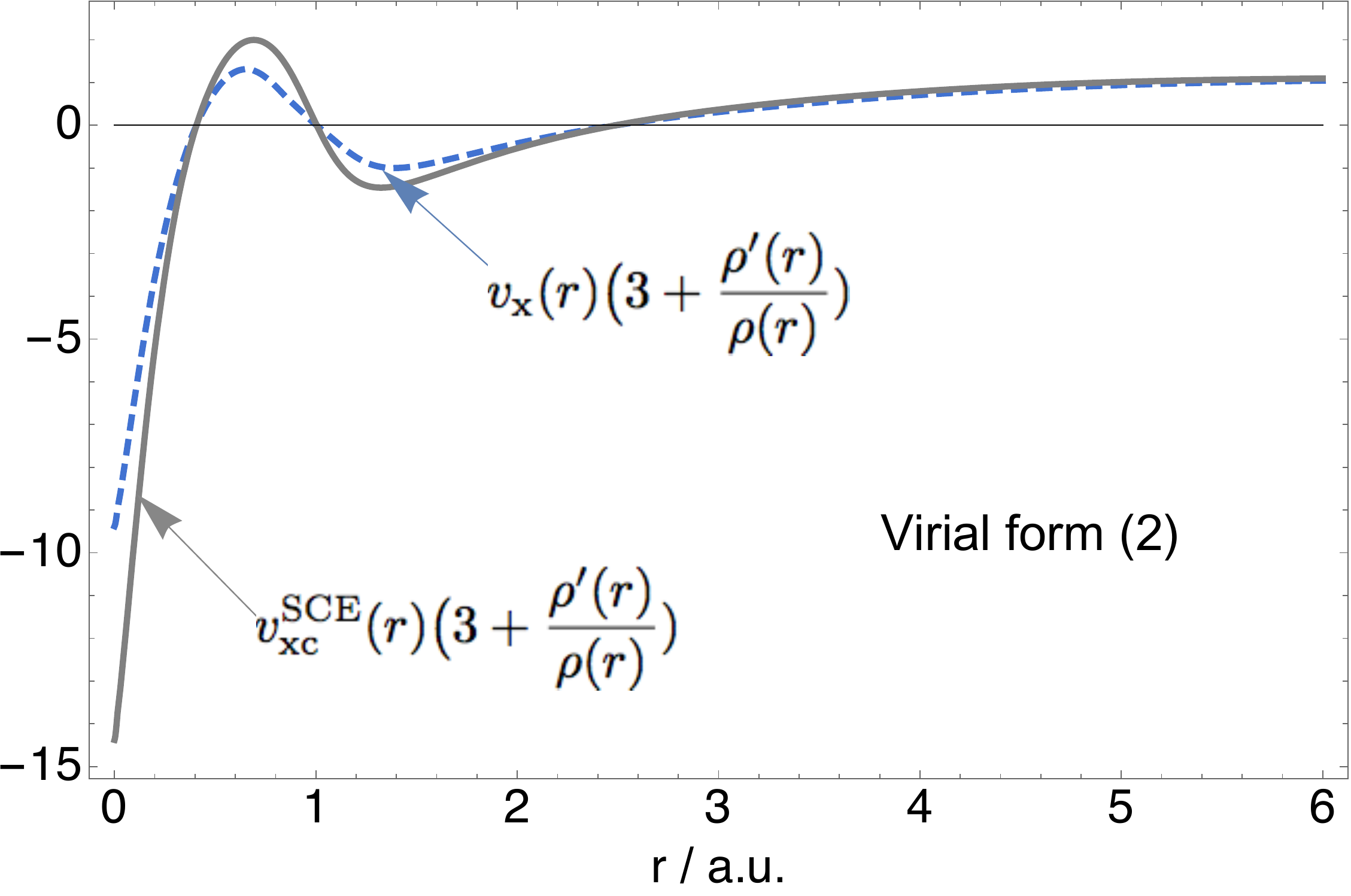}\par
    \includegraphics[width=0.8\linewidth]{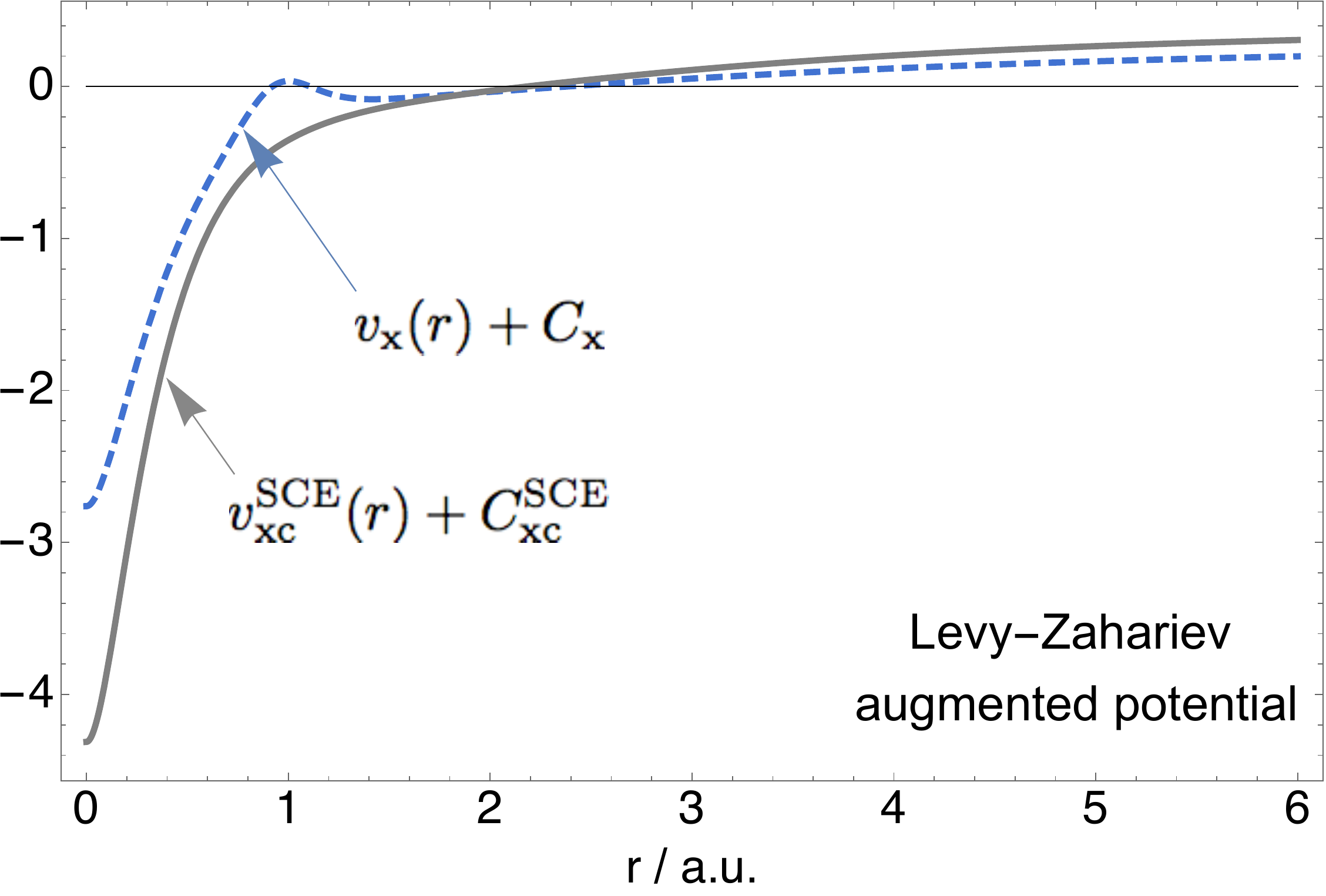}\par
    \includegraphics[width=0.8\linewidth]{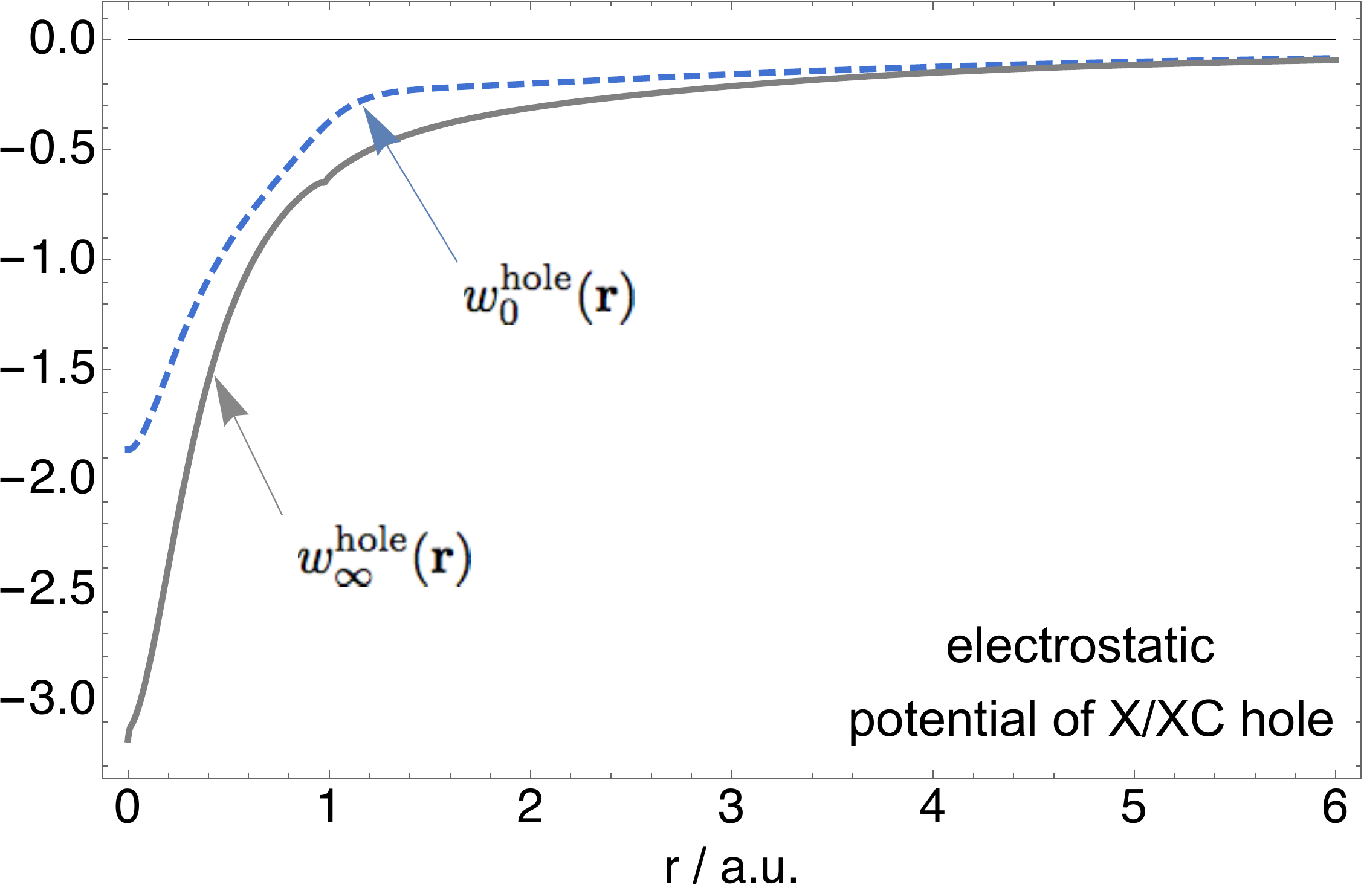}\par
\caption{Weak and strong coupling limit energy densities for the beryllium atom, within the different energy density gauges of Table~\ref{table_edens}.}
\label{fig_edens_be}
\end{figure}

\section{Virial energy densities arising from the exchange and SCE potentials}
\label{sec_virial}
The fact that the exchange and the SCE functional and their potentials have the same behavior under uniform coordinate scaling can be used to obtain useful expressions for these quantities. For example, the fact that the scaling of $E_x[\rho_\gamma]$ and $v_x([\rho_\gamma],\rv)$ is the same as $V_{ee}^{\rm SCE}[\rho_\gamma]$ and $v_{\rm SCE}([\rho_\gamma],\rv)$,\cite{YanLev-PRL-90} respectively, dictates that $C_x[\rho]$ also scales with $\gamma$ and that the augmented exchange potential obeys: 
\begin{align}
v^{\rm LZ}_x([\rho_\gamma],\rv)=\gamma\, v^{\rm LZ}_x([\rho],\gamma \rv).
  \label{eq:scalxpot}
	\end{align}
We also show here that the well known Levy-Perdew virial relation\cite{LevPer-PRA-85} for the exchange potential,
\begin{align}
E_x[\rho]= -\int \rho(\rv) \, \rv \cdot \nabla v_{\rm x}(\rv) \mathrm{d}\, \rv,
  \label{eq:nablavex}
	\end{align}
holds also for the SCE potential,
\begin{align}
V_{ee}^{\rm SCE}[\rho]= -\int \rho(\rv) \,\rv \cdot \nabla v_{\rm SCE}(\rv) \,\mathrm{d} \rv.
  \label{eq:nablavsce}
	\end{align}
The scaling proof for Eq.~\eqref{eq:nablavsce} is the same as that for Eq.~\eqref{eq:nablavex} (see Ref.~\citenum{Lev-DFT-95}). It is however also interesting to see how Eq.~\eqref{eq:nablavsce} arises directly from the mathematical properties of the co-motion functions and from Eq.~\eqref{eq:pot_sce}, as this can teach us how to build approximations that satisfy Eq.~\eqref{eq:nablavsce} by construction. We thus consider first the $N=2$ case: combining Eqs~\eqref{eq:nablavsce} and~\eqref{eq:pot_sce} for $N=2$, we have:
\begin{align}
- \int \rho(\rv) ~ \rv \cdot \nabla v_{\rm SCE}(\rv) \mathrm{d} \rv = \int \rho(\rv) ~ \rv \cdot \frac{\rv-\fv(\rv)}{\left | \rv-\fv(\rv) \right|^3}
  \label{eq:nablav2}
	\end{align}
Adding and subtracting from the right-hand side of the above equation the scalar product with $\fv(\rv)$, we obtain:
\begin{align}\label{eq:nablav2ref}
\int \rho(\rv) ~ \rv \cdot \frac{\rv-\fv(\rv)}{\left | \rv-\fv(\rv) \right|^3}\mathrm{d} \rv &= \int \rho(\rv) \frac{1}{\left | \rv-\fv(\rv) \right|} \rm{d} \rv \\ \nonumber
&+ \int \rho(\rv) ~\fv(\rv) \cdot \frac{\rv-\fv(\rv)}{\left | \rv-\fv(\rv) \right|^3} \rm{d} \rv.
  	\end{align}
By making the change of variables $\mathbf{u}=\fv(\rv)$, i.e. $\rv = \fv^{-1}(\mathbf{u})=\fv(\mathbf{u})$ (for $N=2$ by virtue of the group properties of Eq.~\eqref{eq:groupprop} $\fv(\rv)$ must be its own inverse\cite{Sei-PRA-99,SeiGorSav-PRA-07}), we can rewrite the last term of Eq.~\eqref{eq:nablav2ref} as
\begin{align} \label{eq:nablav2u}
& \int \rho(\rv) ~  \fv(\rv) \cdot \frac{\rv-\fv(\rv)}{\left | \rv-\fv(\rv) \right|^3} \mathrm{d} \rv = \\ \nonumber
& \int J(\fv(\mathbf{u})) \rho\big(\fv(\mathbf{u})\big)~\mathbf{u} \cdot  \frac{\fv(\mathbf{u})-\mathbf{u}}{\left | \fv(\mathbf{u})-\mathbf{u} \right|^3} \rm{d} \mathbf{u}
  	\end{align}
where $J(\fv)$ is the Jacobian of the transformation $\mathbf{u}=\fv(\rv)$. 
Using Eq.~\eqref{eq:difcmf} and Eq.~\eqref{eq:nablav2u} we can further obtain
\begin{align}\label{eq:nablav2uc}
& \int \rho(\rv) ~  \fv(\rv) \cdot \frac{\rv-\fv(\rv)}{\left | \rv-\fv(\rv) \right|^3} \mathrm{d} \rv = \\ \nonumber
&- \int \rho(\rv) ~ \rv \cdot \frac{\rv-\fv(\rv)}{\left | \rv-\fv(\rv) \right|^3} \mathrm{d} \rv.
  	\end{align}
Combining Eqs.~\eqref{eq:nablav2ref} and~\eqref{eq:nablav2uc} we get
\begin{align}
2 \int \rho(\rv) ~ \rv \cdot \frac{\rv-\fv(\rv)}{\left | \rv-\fv(\rv) \right|^3} \mathrm{d} \rv = \int \rho(\rv) \frac{1}{\left | \rv-\fv(\rv) \right|} \mathrm{d} \rv,
  \label{eq:nablav2final}
	\end{align}
which implies exactly Eq.~\eqref{eq:nablavsce} for $N=2$. For many electrons the proof is essentially the same. Considering Eq.~\eqref{eq:pot_sce}, we have:
\begin{align} \label{eq:nablavmany}
&-\int \rho(\rv) ~ \rv \cdot \nabla v_{\rm SCE}(\rv) \mathrm{d} \rv = \int \rho(\rv) ~  \rv \cdot \frac{\rv-\fv_2(\rv)}{\left | \rv-\fv_2(\rv) \right|^3} \mathrm{d} \rv \\ \nonumber
&+  \int \rho(\rv) ~  \rv \cdot \frac{\rv-\fv_3(\rv)}{\left | \rv-\fv_3(\rv) \right|^3} \mathrm{d} \rv +...
 \end{align}
Now we just have to add and subtract from each of the integrals appearing on the right-hand side of Eq.~\eqref{eq:nablavmany}, the scalar product with its own $\fv_i(\rv)$ and repeat the same steps done for the case $N = 2$. The only difference is that now the inverse function in the change of variables will be one of the other co-motion functions by virtue of the group properties  of Eq.~\eqref{eq:groupprop}, but after summation of all the terms the result is the same as for $N=2$. This proof extends to general geometry and general number of particles $N$ the proof of Seidl\cite{Sei-PRA-99} for the case of spherically symmetric systems with $N=2$ electrons.

The Levy-Perdew relation also holds for the xc part of the SCE potential,
\begin{align}
W_\infty[\rho]= -\int \rho(\rv) \,\rv \cdot \nabla v_{\rm xc}^{\rm SCE}(\rv)\, \mathrm{d} \rv,
  \label{eq:nablavscexc}
	\end{align}
since the Hartree functional satisfies it.

In analogy to the alternative form of Eq.~\eqref{eq:nablavex}, which was observed by Engel and Vosko,\cite{EngVos-PRB-93} we  also have the following alternative form of Eq.~\eqref{eq:nablavsce}:
	\begin{align}
V_{ee}^{\rm SCE}[\rho]= \int  v_{\rm SCE}(\rv) \big[3 \rho(\rv) +\rv \cdot \nabla \rho(\rv) \big] \mathrm{d} \rv,
  \label{eq:vscenablarho}
	\end{align}
and the following alternative form of Eq.~\eqref{eq:nablavscexc}:
	\begin{align}
W_\infty[\rho]= \int  v_{\rm xc}^{\rm SCE}(\rv) \big[3 \rho(\rv) +\rv \cdot \nabla \rho(\rv) \big] \mathrm{d} \rv.
  \label{eq:vscexcnablarho}
	\end{align}

\section{Energy densities in the weak and strong coupling limits for small atoms}
\label{sec_edens}
%%%%%%%%%%%%%%%%%
\begin{table*}
\centering
\caption{Mathematical forms of weak and strong coupling limit energy density definitions plotted in Figure~\ref{fig_edens_be}}
\label{table_edens}
%\begin{tabular*}{0.9\textwidth}{l@{\extracolsep{\fill}}mmmmc} \hline\hline\noalign{\vskip 1ex}
\begin{tabular*}{0.9\textwidth}{l@{\extracolsep{\fill}}mmmmc} \hline\hline\noalign{\vskip 1ex}
Definition (gauge) of the energy density &  \epsilon_x(\rv)             & \epsilon_\infty(\rv)  \\ 
Virial form (1)\cite{LevPer-PRA-85,Lev-DFT-95}       & -\rv \cdot \nabla v_{\rm x}(\rv) &-\rv \cdot \nabla v_{\rm xc}^{\rm SCE}(\rv) \\
Virial form (2)\cite{LevPer-PRA-85,EngVos-PRB-93,Lev-DFT-95} & v_x(\rv) \big[3 +\rv \cdot \big(\nabla \rho(\rv)/\rho(\rv)\big) \big]   & v_{xc}^{\rm SCE}(\rv) \big[3 +\rv \cdot \big(\nabla \rho(\rv)/\rho(\rv)\big) \big] \\
Levy and Zahariev augmented potential\cite{LevZah-PRL-14}    & v_{\rm  x}(\rv)+C_{\rm x}[\rho]  & v_{xc}^{\rm  SCE}(\rv)+C_{xc}^{\rm  SCE}[\rho]  \\
Electrostatic potential of the x/xc hole\cite{MirSeiGor-JCTC-12} & \displaystyle \half\int \frac{h_{\rm x}(\mathbf{r},\mathbf{r}')}{|\mathbf{r} - \mathbf{r}'|}\mathrm{d} \rv' & \displaystyle \sum_{k=2}^{N}\frac{1}{2|\rv-\fv_k(\rv)|} - \frac{1}{2} v_{\rm H}(\rv)
\end{tabular*}
\end{table*}
%%%%%%%%%%%%%%%%%
Interpolation along the adiabatic connection between the weak and strong coupling limit is a way of constructing approximate functionals, in which bias towards a particular correlation regime is avoided.\cite{SeiPerLev-PRA-99,SeiPerKur-PRL-00,locpaper,VucGor-JPCL-17} The first attempts in this sense\cite{SeiPerLev-PRA-99,SeiPerKur-PRL-00,GorVigSei-JCTC-09} proposed to interpolate using global (i.e., integrated over all space) quantities, $W_0[\rho]=E_x[\rho]$ (the exchange) and $W_\infty[\rho]$.\cite{SeiPerLev-PRA-99,SeiPerKur-PRL-00} However, interpolation based on local (i.e., energy densities) instead of global quantities is generally more accurate and more amenable to the construction of size-consistent methods.\cite{locpaper,VucIroWagTeaGor-PCCP-17,VucGor-JPCL-17} We can write $W_0[\rho]$ and $W_\infty[\rho]$ in terms of general energy densities $\epsilon_x(\rv)$ and $\epsilon_\infty(\rv)$,
		\begin{align}
W_0[\rho]= \int \epsilon_x(\rv) \rho (\rv) {\rm d} \rv, \qquad W_\infty[\rho]= \int \epsilon_\infty(\rv) \rho (\rv) {\rm d} \rv,
  \label{eq:global2}
	\end{align}
In Figures~\ref{fig_edens_he} and~\ref{fig_edens_be} we compare $\epsilon_x(\rv)$ and $\epsilon_\infty(\rv)$ for the helium and beryllium atoms, respectively, obtained with the different gauges considered in this work. The employed gauges are summarised in Table~\ref{table_edens} and in addition to the conventional gauge, they include the two virial energy densities and the LZ augmented potentials. The computational details are the same as those of Sec.~\ref{sec_atoms} From Figures~\ref{fig_edens_he} and~\ref{fig_edens_be} we can see that the exchange energy density curves, as well as the SCE energy density curves obtained within different gauges have very different structure and shape. We see that only in the case of the gauge of the electrostatic potential of the xc hole $\epsilon_x(\rv)$ lies always above $\epsilon_\infty(\rv)$ (see also Ref.~\onlinecite{MirSeiGor-JCTC-12}). These features make this gauge more suited for the local interpolation schemes, confirming that the choice made in Refs.~\onlinecite{locpaper,ZhoBahErn-JCP-15,BahZhoErn-JCP-16,VucIroWagTeaGor-PCCP-17,VucGor-JPCL-17} is sensible. Moreover, we should not forget that the virial gauges have the major drawback of being origin-dependent.\cite{CruLamBur-JPCA-98}  The LZ gauge, as it could have been predicted from the data of Table~\ref{tab_atoms}, does not provide a clear trend between $\lambda=0$ and $\lambda\to\infty$.
Of course in this work we do not exhaust all the possible choices, see also Ref.~\onlinecite{CruLamBur-JPCA-98} for an in-depth discussion.

\begin{figure}
\includegraphics[width=0.8\linewidth]{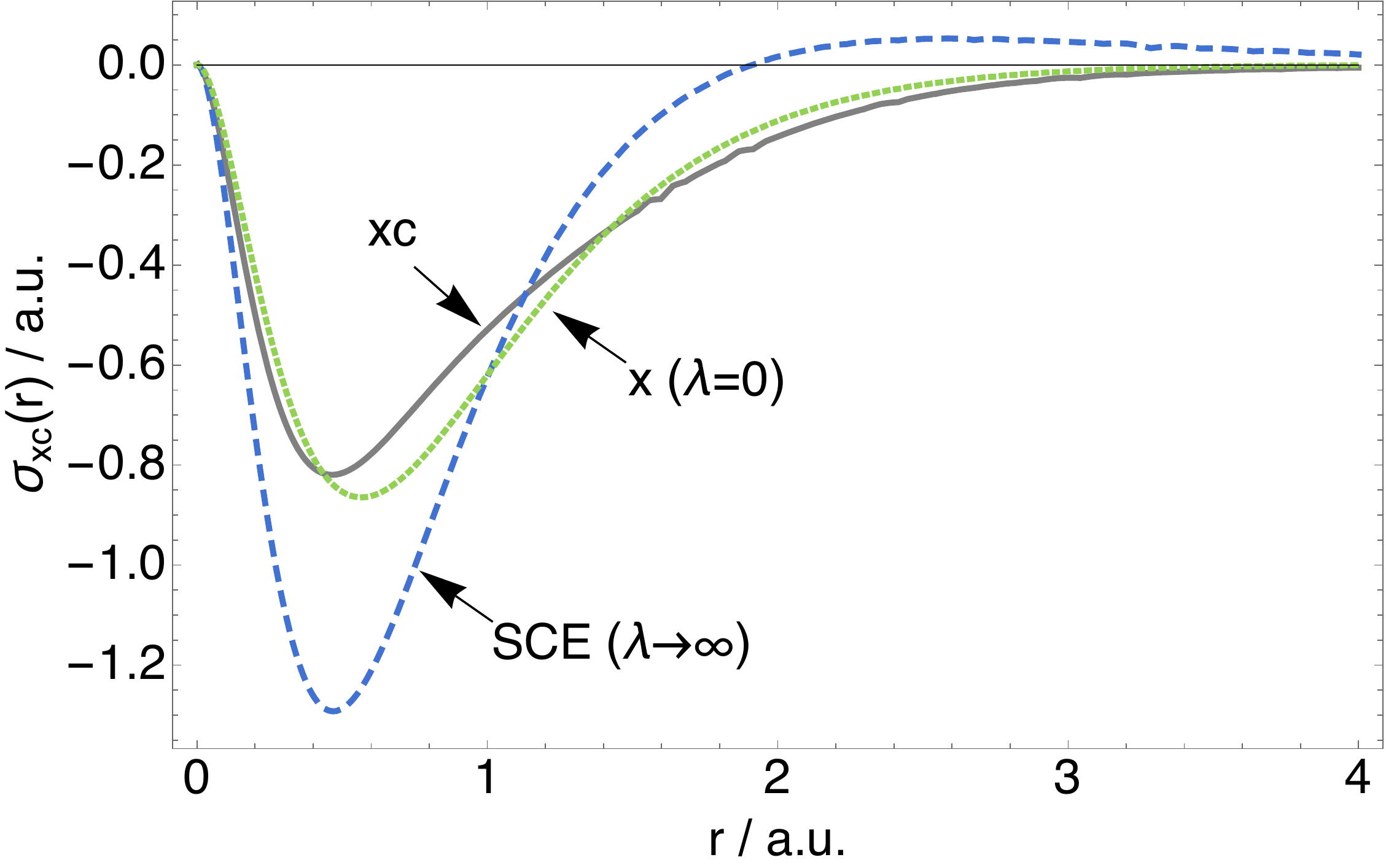}\par
\includegraphics[width=0.8\linewidth]{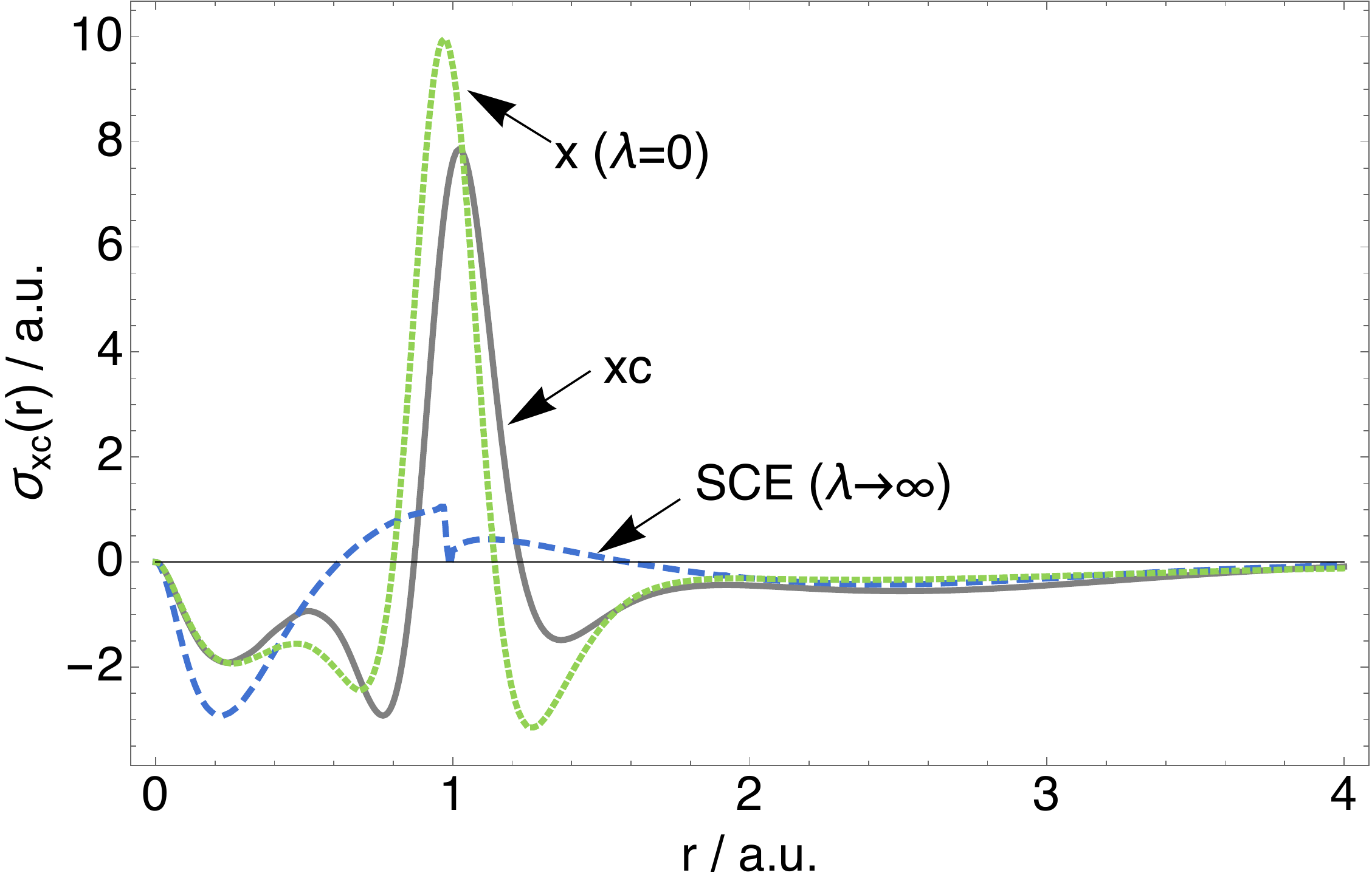}\par
\caption{Plots comparing $\sigma_{xc}(\rv)$ of Eq.~\eqref{eq:sigma_xc} arising from the $v_x(\rv)$, $v_{\rm xc}(\rv)$ and $v_{\rm xc}^{\rm SCE}(\rv)$ potentials for the helium (upper panel) and beryllium (lower panel) atoms}
\label{fig_sigmas}
\end{figure}

\section{Weak and strong coupling limit of the charge associated to the xc potential}
\label{sec_charge}
In addition to the energy density definitions of Table~\ref{table_edens}, in Refs.~\onlinecite{XavGuz-PRL-11,GidLat-JCP-12} it has been proposed that approximations to the xc potential can be built by modelling the {\it fictitious charge} $\sigma_{xc}(\rv)$, which generates the xc potential
\begin{align}
v_{xc} (\rv)=\int \frac{\sigma_{xc}(\rv)}{\left | \rv-\rv' \right|}\,{\rm d}\rv',
\label{eq:sigma_xc1}
\end{align}
or
\begin{align}
\sigma_{xc}(\rv)=- \frac{\nabla^2 v_{xc} (\rv)}{4 \pi}.
\label{eq:sigma_xc}
\end{align}
The main advantage of approximating $v_{\rm xc} (\rv)$ by modelling $\sigma_{\rm xc}(\rv)$ is the fact that, if a model for $\sigma_{\rm xc}(\rv)$ integrates to $-1$ and vanishes at infinity, then the corresponding $v_{xc}(\rv)$ will have the correct asymptotic behaviour $v_{xc}(|\rv|\to\infty)\to -1/r$, which is usually missed by the standard density functional approximations.\cite{XavGuz-PRL-11,GidLat-JCP-12}

In this section we compare the $\sigma_{\rm xc}(\rv)$ fictitious charge arising from the xc potential at physical coupling strength ($\lambda=1$) with $\sigma_{x}(\rv)$ arising from the exchange potential ($\lambda=0$) and $\sigma_{\rm xc}(\rv)$ arising from the $v_{xc}^{\rm SCE}(\rv)$ in the strong coupling limit ($\lambda \to \infty$). In Figure~\ref{fig_sigmas} we show these quantities for the helium and beryllium atoms, using again the accurate potentials described in Sec.~\ref{sec_atoms}. We can see that, similarly to the constant $C[\rho]$, there is no specific trend, and that the effective charge in the SCE limit is quite different than the physical one, except in the valence region of the Be atom. Overall, the only promising way to use the SCE input to construct local interpolation schemes seems to be the use of the energy density in the gauge of the xc-hole electrostatic potential. An even more promising alternative is to use and rescale the SCE mathematical structure to desing approximations for the physical $\lambda=1$ case, as recently proposed in Ref.~\onlinecite{VucGor-JPCL-17}.

\section{Concluding Thoughts}
\label{sec_concl}

In this work we have focused on the exchange-correlation energy densities from the weak and strong coupling limits within different definitions (gauges). In addition to the conventional DFT gauge, which arises directly from a many-body wavefunction via the exchange-correlation hole, we considered other gauges linked to the exchange-correlation potentials, namely the augmented potential of Levy and Zahariev and the virial gauges.
We have also further investigated the features of the augmented LZ potential in the strong coupling limit, which arises very naturally and it is equal to the Kantorovich potential. The LZ shift can be also used to analyze the thermodynamic limit of the classical uniform electron gas, as explained in Sec.~\ref{sec:parlam}.

We have shown that the Kantorovich potential and the augmented exchange potential obey the same simple relation under uniform coordinate scaling, as summarised in Eqs.~\eqref{eq:scalCsceU} and ~\eqref{eq:scalxpot}. We have also shown that the xc part of the SCE potential also obeys the Levy-Perdew virial relation, Eq.~\eqref{eq:nablavscexc}, and thus also the Engel-Vosko relation, Eq.~\eqref{eq:vscexcnablarho}, which is a transformation of Eq.~\eqref{eq:nablavscexc}. These expressions have been used to compare the strong and weak coupling limit energy densities within different gauges for the helium and beryllium atoms (see Figs.~\ref{fig_edens_he} and~\ref{fig_edens_be}), comparing them to the gauge defined by the LZ potential. We have found that only in the case of the gauge of the electrostatic potential of the xc hole the weak and strong coupling limit energy densities do not cross. This observation is important for approaches that model the xc functional by interpolating between the weak and strong coupling limit energy densities.\cite{locpaper,ZhoBahErn-JCP-15,BahZhoErn-JCP-16,VucIroWagTeaGor-PCCP-17,VucGor-JPCL-17,VucGor-JPCL-17} We have also carried out in Sec.~\ref{sec_charge} a similar analysis for the effective charge associated to the xc potential defined in Refs.~\onlinecite{XavGuz-PRL-11,GidLat-JCP-12}, observing that there is no clear trend as the interaction strength increases. In future works it might be useful to also analyze energy densities defined in terms of the modulus square of the electric field.\cite{DeG-priv,Eic-priv}

\section*{Acknowledgements}
We thank Sara Giarrusso for a critical reading of the manuscript and suggestions to improve it. This work was supported by the Netherlands Organization for Scientific Research (NWO) through an ECHO grant (717.013.004) and the European Research Council under H2020/ERC Consolidator Grant corr-DFT (Grant No. 648932). 

\bibliography{biblioPaola,biblio_spec,biblio1,biblio_add,biblio_add2}

\end{document}